\documentclass[letter]{jpsj2}
\usepackage{graphicx}
\usepackage{latexsym}
\usepackage{amsmath}
\usepackage{color}
\newcommand{\vek}[1]{\mbox{\boldmath${#1}$}}
\newcommand{\dis}{\displaystyle}

\def\ba{\begin{align}}

\title{%
Homochirality proliferation in space
}

\author{%
Yukio \textsc{Saito}\thanks{yukio@rk.phys.keio.ac.jp}
 and Hiroyuki \textsc{Hyuga}\thanks{hyuga@rk.phys.keio.ac.jp}
}

\inst{%
Department of Physics, Keio University, Yokohama 223-8522 
}

\recdate{\today}

\abst{%
To understand the chirality selection in the biological organic system,
a simple lattice model of chemical reaction with
molecular diffusion is proposed and studied by Monte Carlo simulations.
In addition to a simple stochastic process of conversions
between an achiral reactant A and chiral products, R and S molecules,
a nonlinear autocatalysis is incorporated by enhancing the reaction rate from
A to R ( or S), 
when A is surrounded by more than one R ( or S) molecules.
With this nonlinear autocatalysis, the chiral symmetry between R and S is
shown to be broken in a dense system.
In a dilute solution with a chemically inactive solvent, molecular diffusion 
 accomplishes chirality  selection, 
 provided that the initial concentration is higher
 than the critical value.
}

\kword{%
chemical physics, chirality selection, enantiometric excess, 
rate equation, reaction-diffusion system, Monte Carlo simulation,
dynamical chiral symmetry break
}

\begin{document}
\sloppy
\maketitle

Organic molecules often have two possible stereo-structures, i.e. 
a right-handed (R) and a mirror-image left-handed (S) form, 
but those associated with living matter choose only one type:
only L-amino acids and D-sugards.
\cite{bonner88,feringa+99}
There are many studies on the origin of this chirality selection.
\cite{bonner88}
Various mechanisms proposed to cause
asymmetry in the primordial molecular environment
( by chance or 
by external or internal deterministic factors) 
\cite{mason+85,kondepudi+85,meiring87,bada95,bailey+98,feringa+99,hazen+01}
 turned out to be very minute, and therefore it has to be amplified. 

Frank has shown theoretically that an autocatalytic reaction 
of a chemical substance with an antagonistic process
can lead to an amplification of enantiometric excess (ee)
and  to homochirality.\cite{frank53}
Recently, 
amplification of ee was confirmed in the asymmetric autocatalysis
 of pyrimidyl alkanol,
\cite{soai+90,soai+95,sato+01,sato+03}
 and the temporal evolution was explained by the second-order autocatalytic
reaction.\cite{sato+01,sato+03} 
In various other systems such as crystallizations, 
the chiral symmetry breaking is found and discussed extensively.
\cite{kondepudi+01}

In our previous paper\cite{sh+04}, 
we have shown that in addition to the nonlinear autotacalysis
 a recycling process of the reactant introduced by the back reaction 
accomplishes the complete homochirality.
There, however, chemical reaction is analyzed macroscopically
in terms of average concentrations.
Thus, a very important factor is neglected, namely the
spatial distribution of the chemical components;
one cannot understand
how the homochirality is established over the system.
In this paper, we study the chemical reaction of molecules 
in an extended space to understand the proliferation of chirality selection.

\noindent
{\it Model and Elementary Processes}

In order to understand the essentials of the chirality selection, we
propose here a simple model such that the space is restricted to two dimensions
and is devided into a lattice.
Molecules are treated as points moving randomly on the lattice sites.
Double occupancy of a lattice site is forbidden.
There involve four types of molecules in the present minimal model; 
an achiral reactant A, two types of product 
enantiomers R and S, and a solvant in a diluted system. 
As for the chemical reaction, three typical cases are analyzed;
nonautocatalytic, linearly autocatalytic and secondary autocatalytic cases.
The back-reaction is always included in those three cases.

The non-autocatalytic chemical reaction proceeds on site 
and independently of neighboring molecules,
as illustrated schematically as\\
\begin{center}
\unitlength 0.1in
\begin{picture}(28.90,5.04)(7.90,-6.94)
%
\special{pn 8}%
\special{pa 2376 198}%
\special{pa 2376 686}%
\special{fp}%
%
\special{pn 8}%
\special{pa 2544 190}%
\special{pa 2544 678}%
\special{fp}%
%
\special{pn 8}%
\special{pa 2216 358}%
\special{pa 2704 358}%
\special{fp}%
%
\special{pn 8}%
\special{pa 2200 510}%
\special{pa 2688 510}%
\special{fp}%
\put(24.0800,-4.7300){\makebox(0,0)[lb]{A}}%
%
\special{pn 8}%
\special{pa 3352 206}%
\special{pa 3352 694}%
\special{fp}%
%
\special{pn 8}%
\special{pa 3520 198}%
\special{pa 3520 686}%
\special{fp}%
%
\special{pn 8}%
\special{pa 3192 366}%
\special{pa 3680 366}%
\special{fp}%
%
\special{pn 8}%
\special{pa 3176 518}%
\special{pa 3664 518}%
\special{fp}%
%
\special{pn 8}%
\special{pa 2816 404}%
\special{pa 3080 404}%
\special{fp}%
\special{sh 1}%
\special{pa 3080 404}%
\special{pa 3013 384}%
\special{pa 3027 404}%
\special{pa 3013 424}%
\special{pa 3080 404}%
\special{fp}%
%
\special{pn 8}%
\special{pa 2816 504}%
\special{pa 3080 504}%
\special{fp}%
\special{sh 1}%
\special{pa 2816 504}%
\special{pa 2883 524}%
\special{pa 2869 504}%
\special{pa 2883 484}%
\special{pa 2816 504}%
\special{fp}%
\put(33.8400,-4.9400){\makebox(0,0)[lb]{R}}%
\put(7.9000,-6.5000){\makebox(0,0)[lb]{A\; $\displaystyle\rightleftharpoons$\; S}}%
\put(8.0000,-4.2000){\makebox(0,0)[lb]{A\; $\displaystyle \rightleftharpoons$\; R}}%
\end{picture}%
\end{center}
The reaction process is essentially stochastic, and we simulate it
by the Monte Carlo method. In the mean-field approximation where the
fluctuation is neglected, the process is described by the
rate equation in terms of
the local concentrations $r(\vek i)$, $s(\vek i)$ and $a(\vek i)$ 
of R, S, and A molecules at a site $\vek i$ as
\begin{align}
\left. \frac{dr(\vek i)}{dt} \right|_0 = k_0 a(\vek i) - \lambda r(\vek i),
\nonumber \\
\left. \frac{ds(\vek i)}{dt}\right|_0 = k_0 a(\vek i) - \lambda s(\vek i),
\nonumber \\
\left. \frac{da(\vek i)}{dt} \right|_0= 
- \left( \left. \frac{dr(\vek i)}{dt} \right|_0 + \left. \frac{ds(\vek i)}{dt}\right|_0 
\right)
\end{align}
with a constant production rate $k_0$ and a decay rate $\lambda$.
The rate equation is very powerful in theoretical analysis.
In this non autocatalytic case, for example,
enantiomer concentrations are shown to approach to 
values at a symmetric fixed point 
$r_{\infty}=s_{\infty}= k_0 a_{\infty}/\lambda$, asymptotically.

Linearly autocatalytic reaction is described by the reaction scheme\\
\begin{center}
\unitlength 0.1in
\begin{picture}(28.86,4.96)(12.50,-8.86)
%
\special{pn 8}%
\special{pa 2656 398}%
\special{pa 2656 886}%
\special{fp}%
%
\special{pn 8}%
\special{pa 2824 390}%
\special{pa 2824 878}%
\special{fp}%
%
\special{pn 8}%
\special{pa 2496 558}%
\special{pa 3080 558}%
\special{fp}%
%
\special{pn 8}%
\special{pa 2480 710}%
\special{pa 3072 710}%
\special{fp}%
%
\special{pn 8}%
\special{pa 3168 646}%
\special{pa 3432 646}%
\special{fp}%
\special{sh 1}%
\special{pa 3432 646}%
\special{pa 3365 626}%
\special{pa 3379 646}%
\special{pa 3365 666}%
\special{pa 3432 646}%
\special{fp}%
%
\special{pn 8}%
\special{pa 2992 390}%
\special{pa 2992 878}%
\special{fp}%
%
\special{pn 8}%
\special{pa 3712 398}%
\special{pa 3712 886}%
\special{fp}%
%
\special{pn 8}%
\special{pa 3880 390}%
\special{pa 3880 878}%
\special{fp}%
%
\special{pn 8}%
\special{pa 3552 558}%
\special{pa 4136 558}%
\special{fp}%
%
\special{pn 8}%
\special{pa 3536 710}%
\special{pa 4128 710}%
\special{fp}%
%
\special{pn 8}%
\special{pa 4048 390}%
\special{pa 4048 878}%
\special{fp}%
\put(26.900,-6.8600){\makebox(0,0)[lb]{A}}%
\put(39.1000,-6.9200){\makebox(0,0)[lb]{R}}%
\put(28.6000,-6.8600){\makebox(0,0)[lb]{R}}%
\put(37.500,-6.9200){\makebox(0,0)[lb]{R}}%
\put(5.6000,-6.3000){\makebox(0,0)[lb]{A\;$\displaystyle +$\;R\;$\displaystyle\rightarrow$\;2R}}%
\put(5.5000,-8.9000){\makebox(0,0)[lb]{A\;$\displaystyle +$\;S\;$\displaystyle\rightarrow$\;2S}}%
\end{picture}%
\end{center}
Since the double occupancy of a lattice site is forbidden in the present
lattice model, it is natural to assume
that these autocatalytic reactions take place 
when an A molecule is located next to
one or more R or S molecules, respectively, with a probability $k_1$.
Then the additional contribution to the rate equation (1) is described as
\begin{align}
\left. \frac{dr(\vek i)}{dt} \right|_1 =  k_1 a(\vek i) 
\dis{\sum_{\vek i_1} r(\vek i_1)},
\nonumber \\
\left. \frac{ds(\vek i)}{dt} \right|_1 =  k_1 a(\vek i) 
\dis{\sum_{\vek i_1} s(\vek i_1)},
\nonumber \\
\left. \frac{da(\vek i)}{dt} \right|_1= 
- \left( \left. \frac{dr(\vek i)}{dt} \right|_1 + \left. \frac{ds(\vek i)}{dt}\right|_1 
\right)
\end{align}
where the summation of sites $\vek i_1$ runs over the 4
nearest neighboring ones to $\vek i$.

The nonlinear autocatalysis of the second order is described by the reaction \\
\begin{center}
\unitlength 0.1in
\begin{picture}(30.14,9.76)(5.30,-14.66)
%
\special{pn 8}%
\special{pa 2360 1098}%
\special{pa 2360 1466}%
\special{fp}%
%
\special{pn 8}%
\special{pa 1952 498}%
\special{pa 1952 1018}%
\special{fp}%
%
\special{pn 8}%
\special{pa 2120 490}%
\special{pa 2120 1010}%
\special{fp}%
%
\special{pn 8}%
\special{pa 1840 578}%
\special{pa 2360 578}%
\special{fp}%
%
\special{pn 8}%
\special{pa 1840 730}%
\special{pa 2360 730}%
\special{fp}%
\put(19.9200,-7.0600){\makebox(0,0)[lb]{A}}%
%
\special{pn 8}%
\special{pa 2520 738}%
\special{pa 2784 738}%
\special{fp}%
\special{sh 1}%
\special{pa 2784 738}%
\special{pa 2717 718}%
\special{pa 2731 738}%
\special{pa 2717 758}%
\special{pa 2784 738}%
\special{fp}%
%
\special{pn 8}%
\special{pa 2288 490}%
\special{pa 2288 1010}%
\special{fp}%
\put(21.5000,-7.0600){\makebox(0,0)[lb]{R}}%
%
\special{pn 8}%
\special{pa 1840 898}%
\special{pa 2352 898}%
\special{fp}%
\put(19.9200,-8.7400){\makebox(0,0)[lb]{R}}%
%
\special{pn 8}%
\special{pa 2976 498}%
\special{pa 2976 1010}%
\special{fp}%
%
\special{pn 8}%
\special{pa 3136 498}%
\special{pa 3136 1018}%
\special{fp}%
%
\special{pn 8}%
\special{pa 2888 586}%
\special{pa 3384 586}%
\special{fp}%
%
\special{pn 8}%
\special{pa 2888 738}%
\special{pa 3376 738}%
\special{fp}%
%
\special{pn 8}%
\special{pa 3304 490}%
\special{pa 3304 1010}%
\special{fp}%
\put(31.7000,-7.1400){\makebox(0,0)[lb]{R}}%
%
\special{pn 8}%
\special{pa 2888 906}%
\special{pa 3376 906}%
\special{fp}%
\put(30.1500,-8.8200){\makebox(0,0)[lb]{R}}%
\put(30.1500,-7.1400){\makebox(0,0)[lb]{R}}%
%
\special{pn 8}%
\special{pa 1800 1202}%
\special{pa 2432 1202}%
\special{fp}%
%
\special{pn 8}%
\special{pa 1800 1362}%
\special{pa 2432 1362}%
\special{fp}%
\put(20.6000,-13.3800){\makebox(0,0)[lb]{A}}%
%
\special{pn 8}%
\special{pa 2536 1282}%
\special{pa 2800 1282}%
\special{fp}%
\special{sh 1}%
\special{pa 2800 1282}%
\special{pa 2733 1262}%
\special{pa 2747 1282}%
\special{pa 2733 1302}%
\special{pa 2800 1282}%
\special{fp}%
\put(22.2000,-13.3800){\makebox(0,0)[lb]{R}}%
\put(18.9600,-13.3800){\makebox(0,0)[lb]{R}}%
%
\special{pn 8}%
\special{pa 1864 1098}%
\special{pa 1864 1466}%
\special{fp}%
%
\special{pn 8}%
\special{pa 2912 1194}%
\special{pa 3544 1194}%
\special{fp}%
%
\special{pn 8}%
\special{pa 2912 1354}%
\special{pa 3544 1354}%
\special{fp}%
\put(33.4000,-13.3000){\makebox(0,0)[lb]{R}}%
\put(30.0800,-13.3000){\makebox(0,0)[lb]{R}}%
\put(31.6800,-13.3000){\makebox(0,0)[lb]{R}}%
%
\special{pn 8}%
\special{pa 2016 1090}%
\special{pa 2016 1458}%
\special{fp}%
%
\special{pn 8}%
\special{pa 2184 1098}%
\special{pa 2184 1466}%
\special{fp}%
%
\special{pn 8}%
\special{pa 2976 1098}%
\special{pa 2976 1466}%
\special{fp}%
%
\special{pn 8}%
\special{pa 3128 1098}%
\special{pa 3128 1466}%
\special{fp}%
%
\special{pn 8}%
\special{pa 3296 1098}%
\special{pa 3296 1466}%
\special{fp}%
%
\special{pn 8}%
\special{pa 3464 1098}%
\special{pa 3464 1466}%
\special{fp}%
\put(0.3000,-9.3000){\makebox(0,0)[lb]{A\;$\displaystyle +$\;2R\;$\displaystyle\rightarrow$\;3R}}%
\put(0.3000,-11.5000){\makebox(0,0)[lb]{A\;$\displaystyle +$\;2S\;$\displaystyle\rightarrow$\;3S}}%
\end{picture}%
\end{center}
The situation is achieved in our lattice system by assuming that the reaction
proceeds when an A molecule is surrounded by more than one R or S
molecules. By denoting the corresponding rate as $k_2$, the additional
 contribution to the rate equation (1) is described as
\begin{align}
\left. \frac{dr(\vek i)}{dt} \right|_2 =  k_2 a(\vek i) 
\dis{\sum_{\langle \vek i_1, \vek i_2 \rangle} r(\vek i_1)r(\vek i_2)},
\nonumber \\
\left. \frac{ds(\vek i)}{dt} \right|_2 =  k_2 a(\vek i) 
\dis{\sum_{\langle \vek i_1, \vek i_2 \rangle} s(\vek i_1)s(\vek i_2)},
\nonumber \\
\left. \frac{da(\vek i)}{dt} \right|_2= 
- \left( \left. \frac{dr(\vek i)}{dt} \right|_2 + \left. \frac{ds(\vek i)}{dt}\right|_2 
\right)
\end{align}
where summation over the pairs of sites $\vek i_1$ and $\vek i_2$
runs over 6 combinations of the nearest neighboring sites to $\vek i$.

In addition to the above chemical reaction processes, diffusion of molecules
is important if the reaction system is diluted in chemically inactive solvent.
We assume in the following that the diffusion is essentially mediated by the
solvent molecules.

\noindent
{\it Monte Carlo simulation}

In order to analyze the complex reaction-diffusion system 
in an spatially extended situation, numerical simulation is useful.
We adopt a Monte Carlo simulation in two dimensions
with a following scheme.

As for the initial condition, A molecules with a concentration $c_0$
are distributed randomly on a square lattice of a size $L \times L$, 
and the remaining sites are assumed to be occupied by the solvent.
Periodic boundary conditions are imposed in the $x$ and $y$-directions,
and the length is measured in terms of a lattice constant, hereafter.
Then the Monte Carlo simulation starts.
One selects randomly a site among $L \times L$ square lattice sites.
If it is occupied by an A molecule, one tries the reaction from A 
to R with a probability $k(A \rightarrow R)$ and 
from A to S with a probability $k(A \rightarrow S)$. 
If the A molecule is isolated, $k(A \rightarrow R) = k(A \rightarrow S) =k_0$.
If it is surrounded by one R molecules, 
$k(A \rightarrow R) = k_0+k_1$,
and by more than one R molecules, $k(A \rightarrow R) = k_0+k_1+k_2$.
The similar procedure holds when 
it is surrounded by one or more than one S molecules as
$k(A \rightarrow S) = k_0+k_1$ and $k(A \rightarrow S) = k_0+k_1+k_2$,
respectively.
If the randomly selected site is occupied by an R or S molecule,
it is converted to an A molecule with a probability $\lambda$.

In addition to this reaction algorithm, we can include the diffusion 
process, if necessary. A pair of nearest neighboring sites 
is chosen randomly,
and only if one of them is occupied by a solvent molecule, the
molecules on the chosen pair sites are exchanged. 
We assume no special bonding among A, R and S molecules, and
their diffusion constant is the same, for simplicity.
Their proximity in space only affects the chemical reaction.
Direct exchange among A, R and S molecules is excluded so as 
to realize the diffusionless reaction in the $c_0=1$ case.
One Monte Carlo step (MCS) corresponds to $L \times L$ trials of
 chemical reaction and diffusion, and thus, on average,
each molecule tries reaction and diffusion steps once per each MCS.
The diffusion constant is $D=1/4$ in this time and space unit.

\noindent
{\it Results and analysis for the case without diffusion}

For the chirality selection, the macroscopic rate equation shows that
the nonlinear autocatalysis plays an essential role.
We first perform simulations only with chemical reaction to check if
our simulation model has the chiral symmetry breaking.
The system simulated has a size $L=100$ 
with parameters $k_0=\lambda = 10^{-3}$, and we change the initial 
concentration $c_0$ and the parameters
$k_1$ and $k_2$ in order to find out the role of autocatalytic reaction.

Without diffusion, it is obvious that the result depends strongly 
on the density $c_0$ of the reactive molecules.
Denser the reactive molecules are, more dominates the nonlinearity.
The extreme of the dense system is the one without solvent, $c_0=1$;
a whole lattice sites are initially occupied by A molecules.
At $c_0=1$ without autocatalysis as $k_1=k_2=0$, Monte Carlo simulation shows
that the numbers of R and S molecules  increase synchronously,
and each saturates about one third of the total lattice sites.
The saturation values corresponds to the concentrations at the
fixed point of the rate equation (1);
$r_{\infty}=s_{\infty}=k_0 c_0/(\lambda + 2 k_0)$ and
$a_{\infty}=\lambda c_0/(\lambda + 2 k_0)$,
with the initial concentration $c_0=1$.
The asymptotic spatial configuration (not shown) is completely irregular,
since a molecule on every sites changes its state independent of each other.
The enantiomatic excess, ee, is defined, as usual, 
by the difference in the concentrations of R and S molecules as 
\begin{align}
\phi = \frac{r-s}{r+s}.
\end{align}
Without autocatalysis it fluctuates around zero. 
\begin{figure}[h]
\begin{center} 
\includegraphics[width=0.30\linewidth]{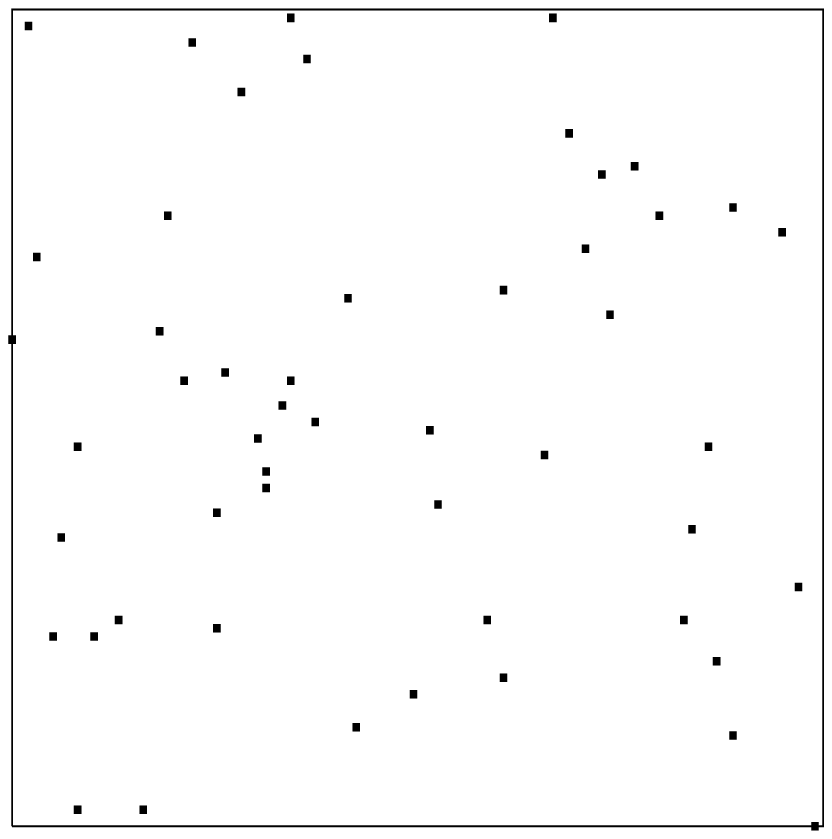}
\includegraphics[width=0.30\linewidth]{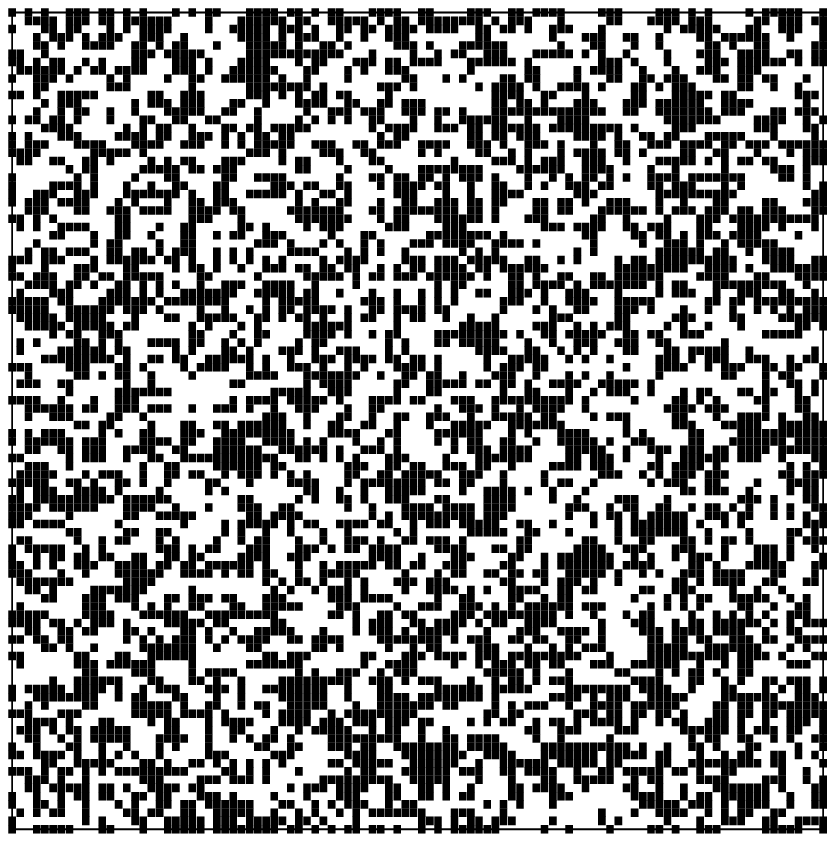}
\includegraphics[width=0.30\linewidth]{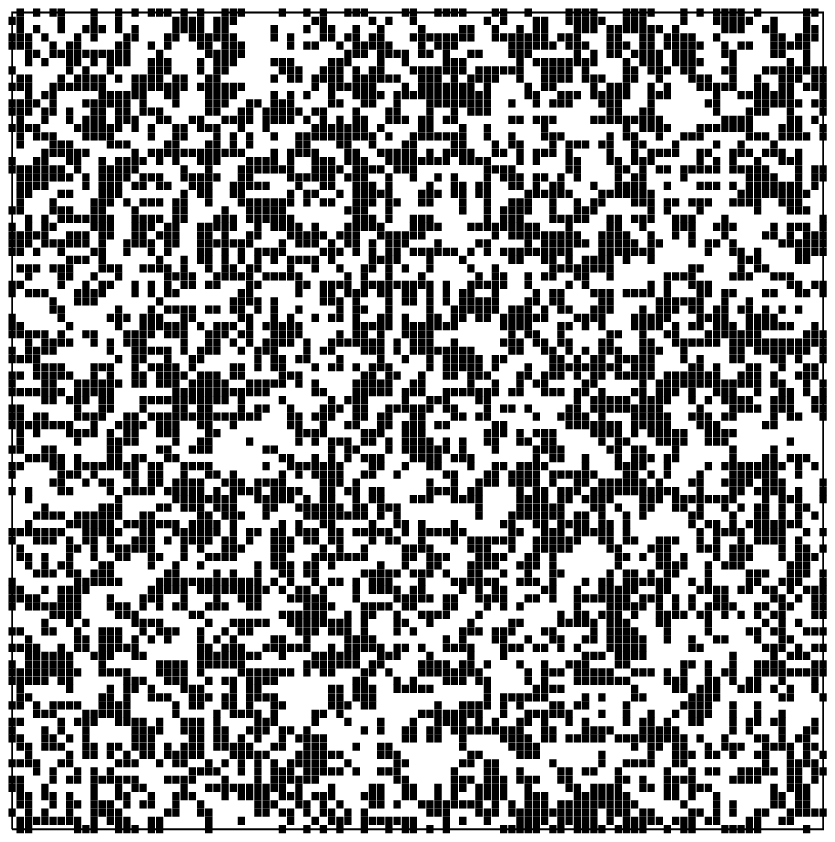}\\
(a) \hspace{4cm} (b) \hspace{4cm} (c)
\end{center} 
\caption{Configurations of (a) A, (b) R, and (c) S molecules
with linear autocatalysis at $10^6$th MCS with $c_0=1$.
}
\label{fig1}
\end{figure}

With a linear autocatalysis as $k_1=100k_0$ but $k_2=0$,
the reaction has produced
 more R and S molecules( in Fig.1(b) and (c), respectively), and less A
molecule ( in Fig.1(a)) in a final configuration,
but the chiral symmetry is not broken; $\phi$ fluctuates around zero,
as shown in Fig.2(d). 
If the spatial dependence is neglected, the corresponding rate equations
are written as
\begin{align}
\frac{dr}{dt}= (k_0+4 k_1r)a-\lambda r   
\nonumber \\
\frac{ds}{dt}= (k_0+ 4k_1s)a-\lambda s   
\end{align}
supplimented with the conservation that the sum of the concentrations 
of A, R and S
molecules is fixed to its initial value $c_0$; $a+r+s=c_0$.
This rate equation has  a symmetric fixed point: 
$r_{\infty}=s_{\infty}=(c_0-a_{\infty})/2
\approx (c_0/2)-(\lambda/8k_1)$ under the assumption of a strong autocatalysis,
 $k_0 \approx \lambda \ll k_1$. With the present parameters,
$c_0=1, ~ k_0=\lambda=k_1/100=10^{-3}$, the asymptotic values 
$r_{\infty}=s_{\infty}=0.499$
are expected, in fair agreement with the simulation result.
Of course, in this case, 
the chiral symmetry should be conserved as the simulation confirms.

\begin{figure}[h]
\begin{center} 
\includegraphics[width=0.30\linewidth]{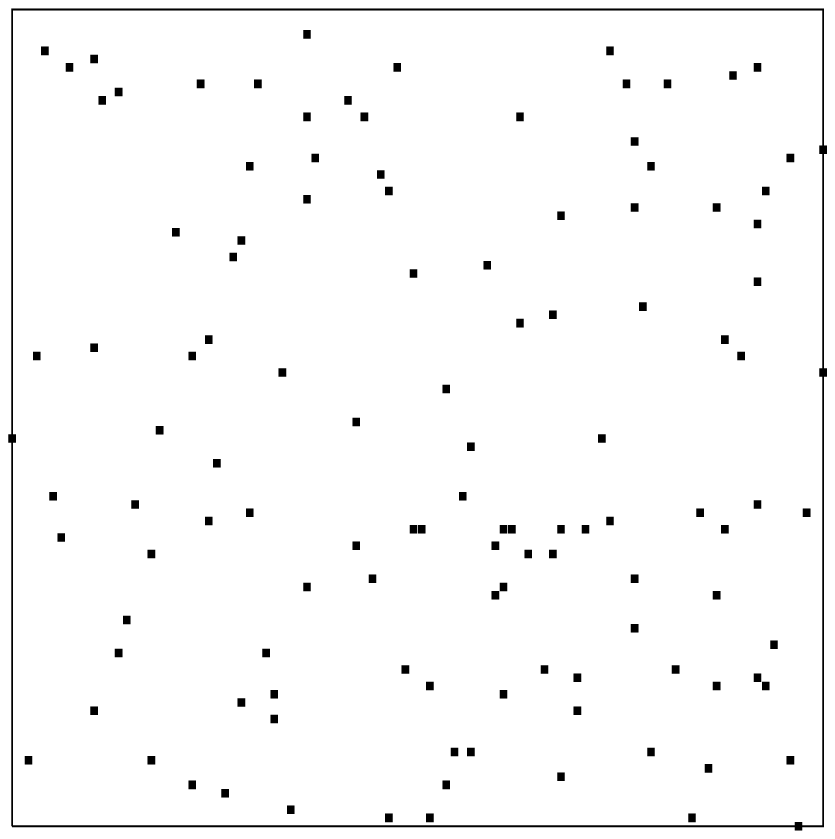}
\includegraphics[width=0.30\linewidth]{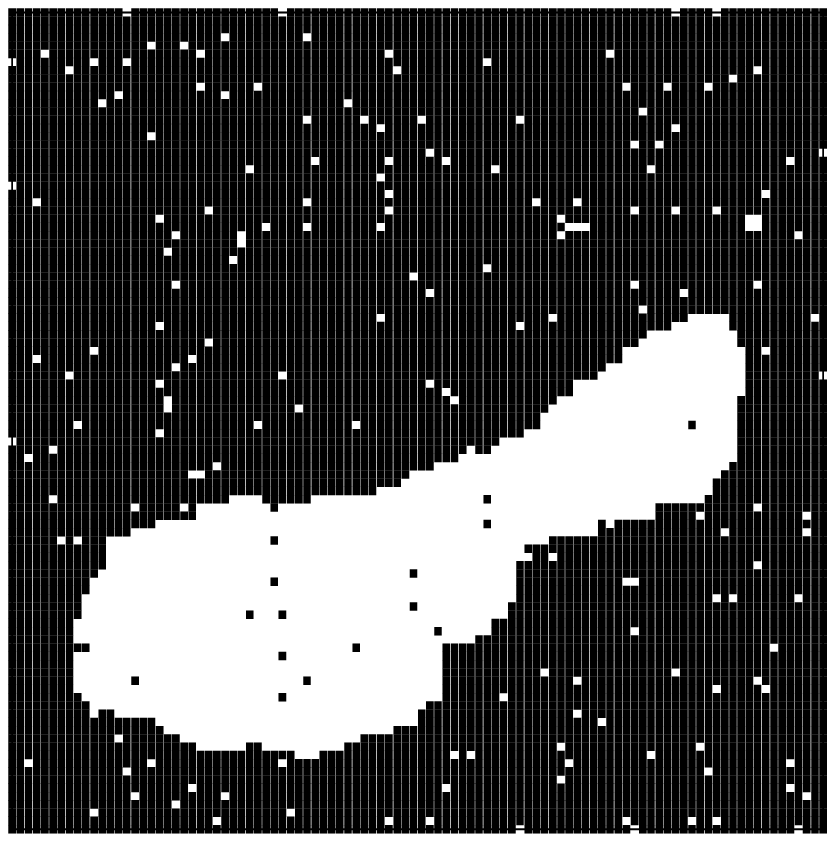}
\includegraphics[width=0.30\linewidth]{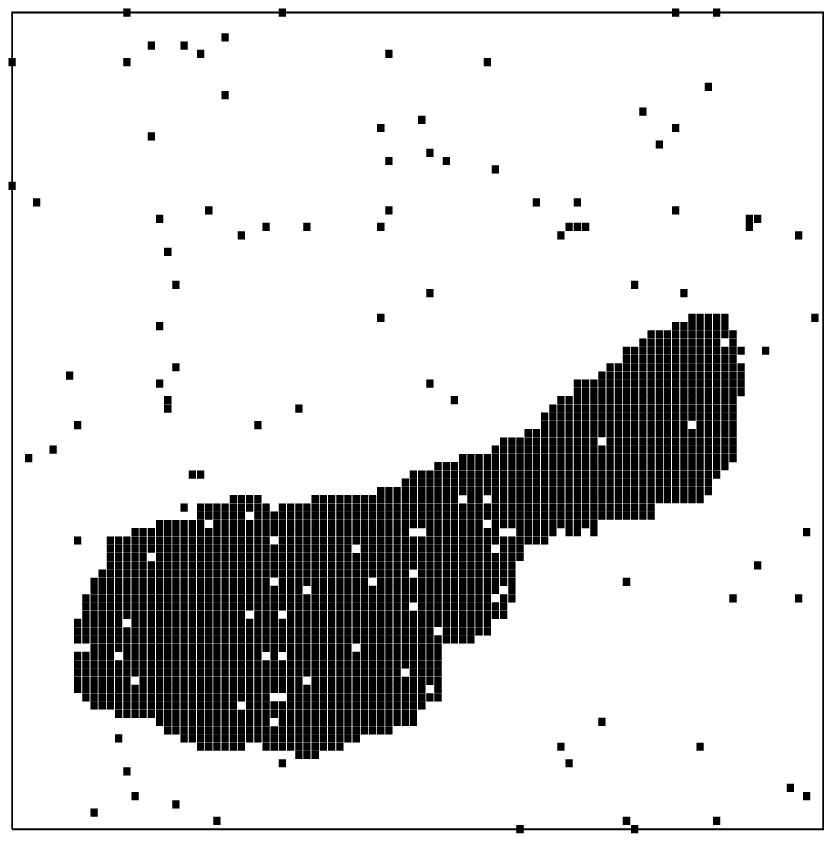}\\
(a) \hspace{4cm} (b) \hspace{4cm} (c)\\
\includegraphics[width=0.36\linewidth]{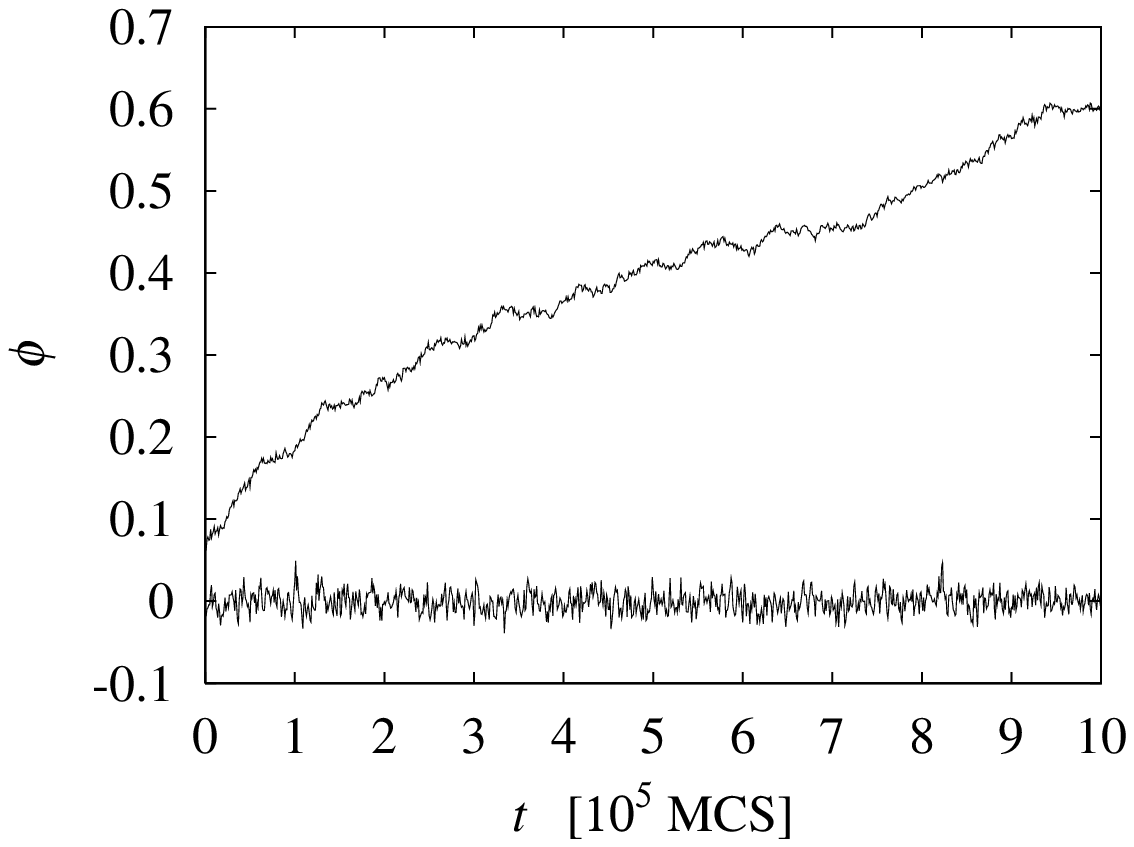}\\
(d)
\end{center} 
\caption{Configurations of (a) A, (b) R, and (c) S molecules
with nonlinear autocatalysis at $10^6$th MCS with $c_0=1$.
(d) Evolution of the enantiometric excess $\phi$.
Upper curve corresponds to the case with nonlinear autocatalysis,
and the lower curve to that with linear autocatalysis, shown in Fig.1.
}
\label{fig2}
\end{figure}

With a second-order autocatalysis with $k_2=100k_0$ but $k_1=0$, 
the chiral symmetry breaks as is shown in Fig.2(a-c).
In the simulation shown, there are more R molecules (Fig.2(b)) 
than S (Fig.2(c)).
By using another sequence of pseudo-random numbers, there
occurs equally cases that the enantiomer S dominates over R.
By neglecting the space dependence the expected rate equations
are written as
\begin{align}
\frac{dr}{dt}= (k_0+ 6 k_2r^2)a-\lambda r   
\nonumber \\
\frac{ds}{dt}= (k_0+ 6k_2s^2)a-\lambda s   
\end{align}
with $a=c_0-r-s$. There is a symmetric fixed point U at 
$r_U=s_U \approx (c_0/2)- (\lambda/6k_2c_0)$ 
and $a_U \approx  \lambda/3 k_2c_0$,
but it is unstable at a high concentration $k_2c_0^2 \gg k_0, ~ \lambda$.
There are stable fixed points at
S$_1$: $(r_{S_1}, s_{S_1},a_{S_1}) \approx 
(c_0-(k_0+\lambda)/6 k_2c_0, ~k_0/6k_2c_0,~ \lambda/6k_2c_0)$
and at 
S$_2$: $(r_{S_2}, s_{S_2},a_{S_2}) \approx 
(k_0/6 k_2c_0, ~c_0-(k_0+\lambda)/6k_2c_0,~ \lambda/6k_2c_0)$.
The amplitude of the ee is expected to approach to the value given by
\begin{align}
|\phi_{\infty}| \approx 1- k_0/3k_2c_0^2,
\end{align}
which is close to unity in the present parameter values, $k_2=100k_0$ and
$c_0=1$.
In the simulation, the ee increases gradually as shown by an upper curve
in Fig.2(d), 
but the final asymptotics is not reached yet.

On looking into the spatial distribution of molecules, one finds that
the sites occupied by R and S molecules form respective domains.
The chirality selection proceeds via the competition between two domains.
The process is very slow and the configuration in Fig.2(a-c)
shows an intermediate stage of the S domain shrinking.
If both R and S domains extend the whole system, the relaxation process
becomes even slower.
The situation looks quite similar
to the slow domain dynamics observed in the spinodal decomposition.

One may notice that the ee in eq.(7) is
complete if the nonautocatalytic production is absent ($k_0=0$),
in agreement with our previous study.\cite{sh+04}
In the present model, we assume a finite value of $k_0$, 
since the creation of an initial chiral molecule R or S from an 
achiral molecule A is necessary. Also without $k_0$, the
accidental extinction of the chiral species, R or S, is unavoidable
and it cannot be recovered only with autocatalytic reactions.

\begin{figure}[h]
\begin{center} 
\includegraphics[width=0.30\linewidth]{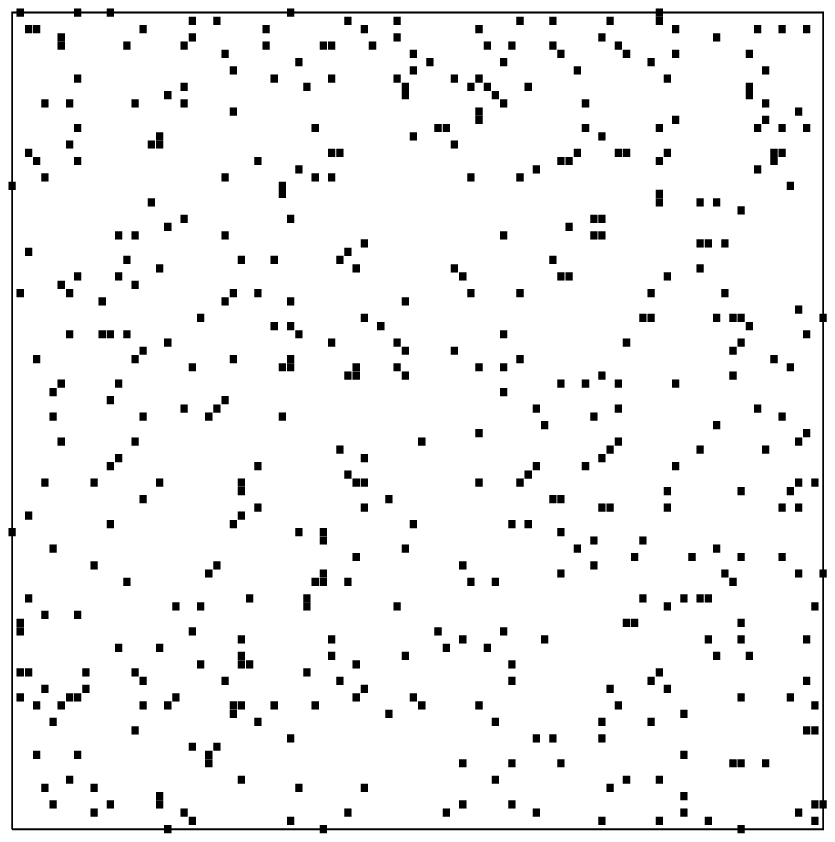}
\includegraphics[width=0.30\linewidth]{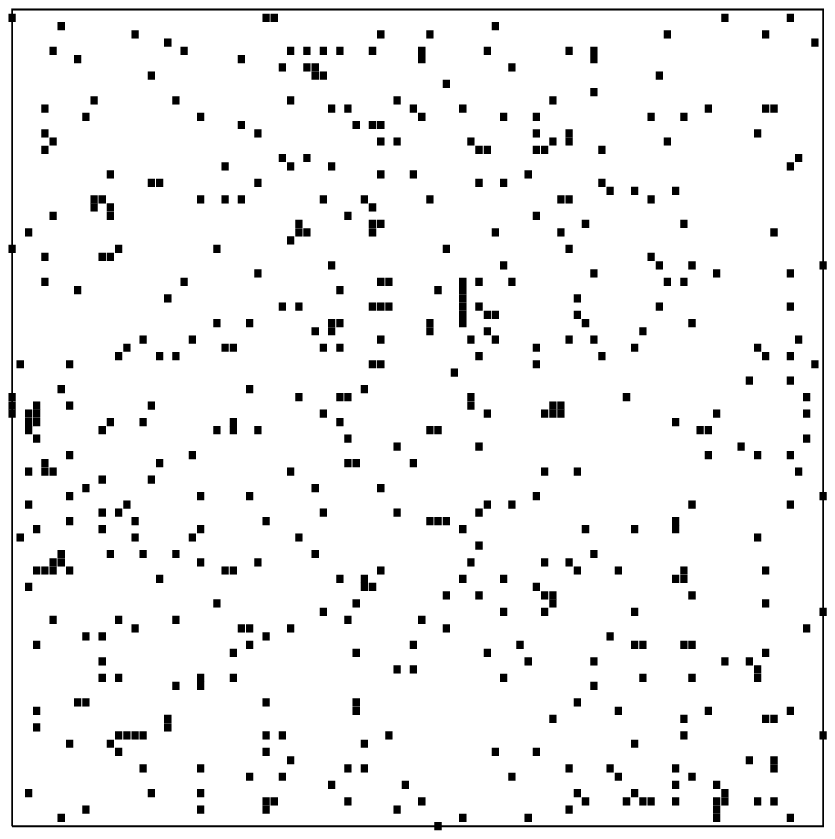}
\includegraphics[width=0.30\linewidth]{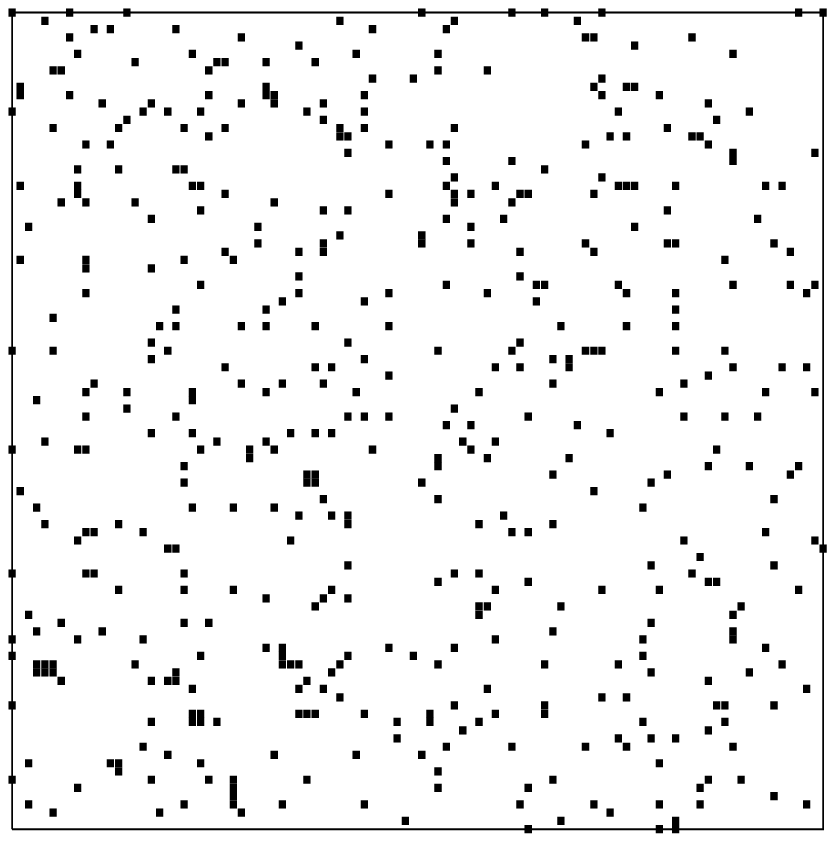}\\
(a) \hspace{4cm} (b) \hspace{4cm} (c)\\
\end{center} 
\caption{Configurations of (a) A, (b) R, and (c) S molecules
with a concentration  $c_0=0.15$ 
at $3 \times 10^4$th MCS  without  diffusion. 
}
\label{fig3}
\end{figure}

We now consider the reaction and chirality selection in a solution with
$c_0 <1$. 
The reaction system is diluted by adding 
inactive solvent molecules. 
If the initial concentration of reactant $c_0$ is
sufficiently high, the nonlinear autocatalysis leads to the chiral symmetry
breaking similar to the case at $c_0=1$. 
On the other hand, at a low concentration
far below the percolation threshold of the square lattice\cite{stauffer85}
$c_{p} \approx 0.6$, the autocatalysis effect cannot propagate through
the whole system and fails to break the chiral symmetry.
For example, at $c_0=0.15$, the numbers of A, R and S molecules 
are about the same with each other as shown in Fig.3(a), (b) and (c),
respectively. There is no cooperative organization, 
and the ee fluctuates around zero, as shown in Fig.4(d).

\begin{figure}[h]
\begin{center} 
\includegraphics[width=0.30\linewidth]{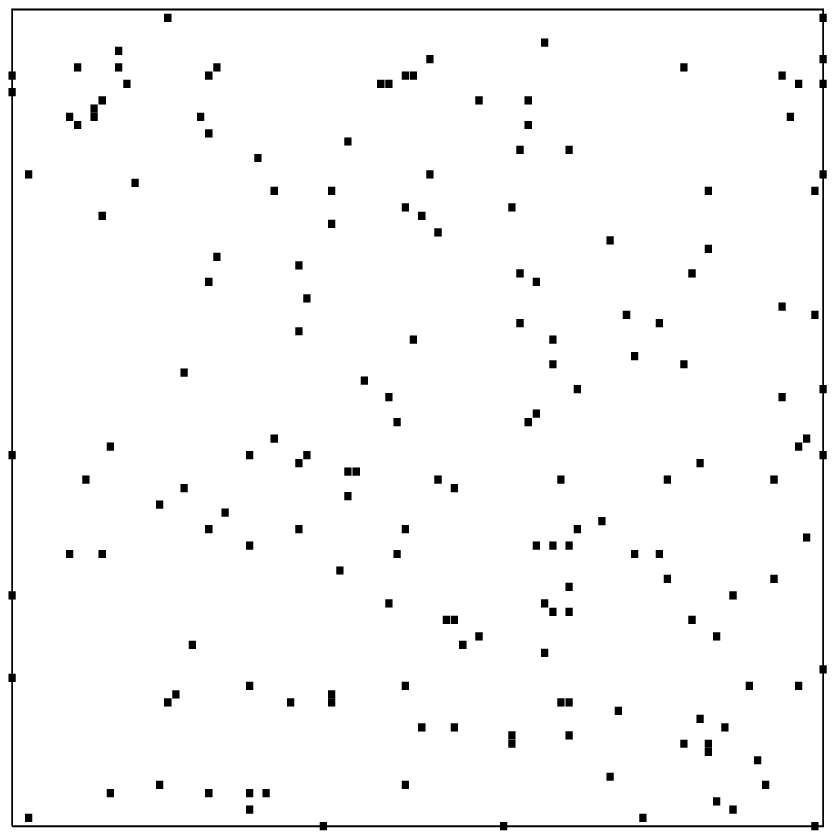}
\includegraphics[width=0.30\linewidth]{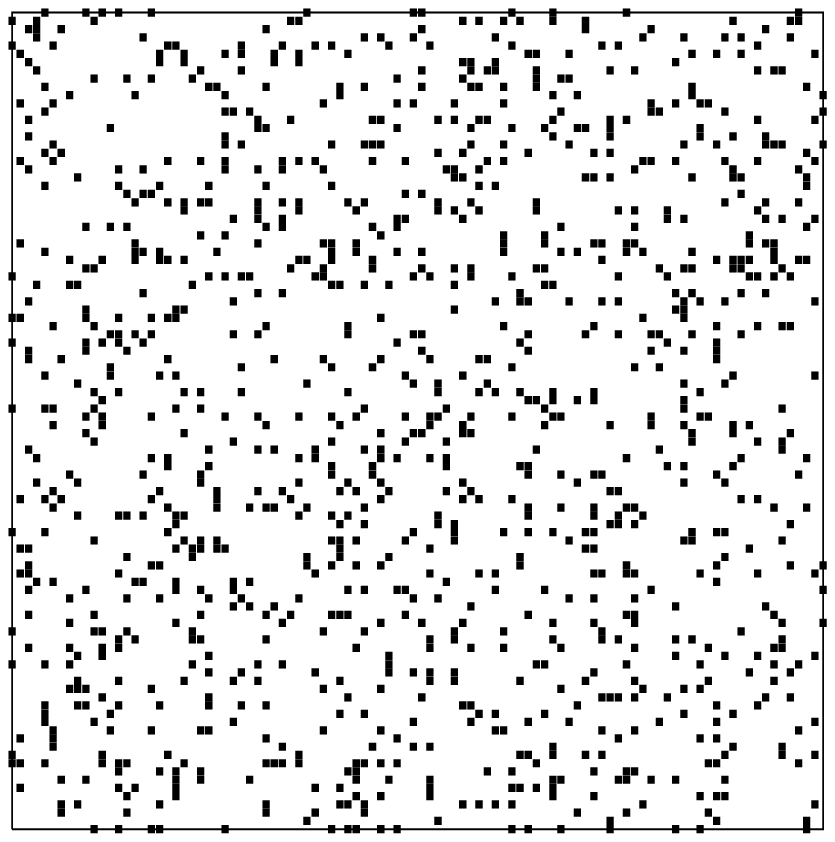}
\includegraphics[width=0.30\linewidth]{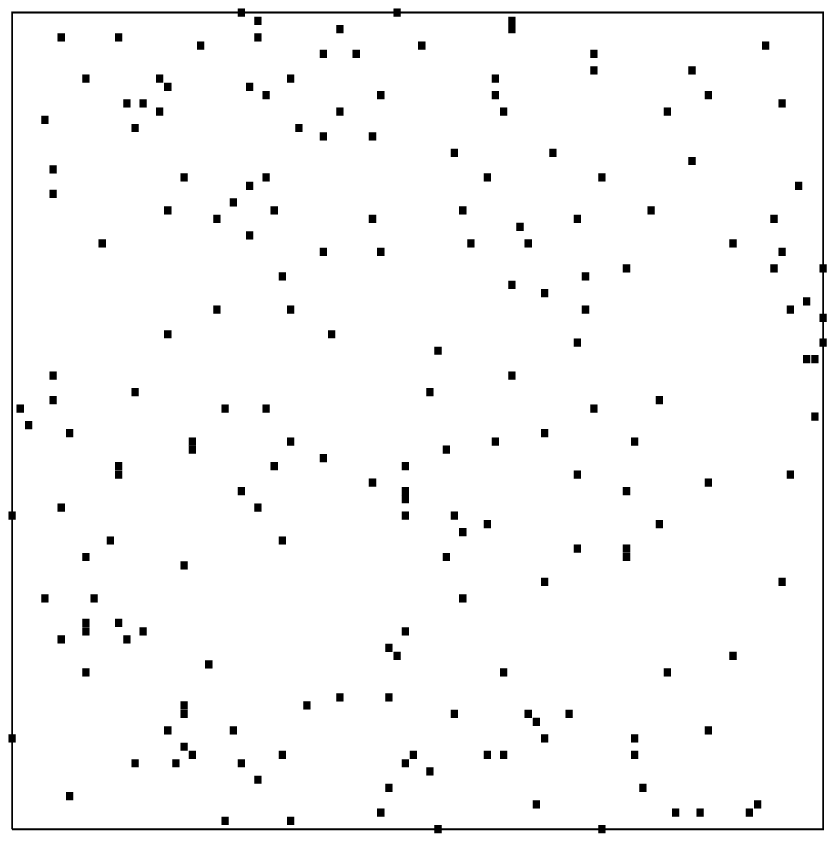}\\
(a) \hspace{4cm} (b) \hspace{4cm} (c)\\
\includegraphics[width=0.40\linewidth]{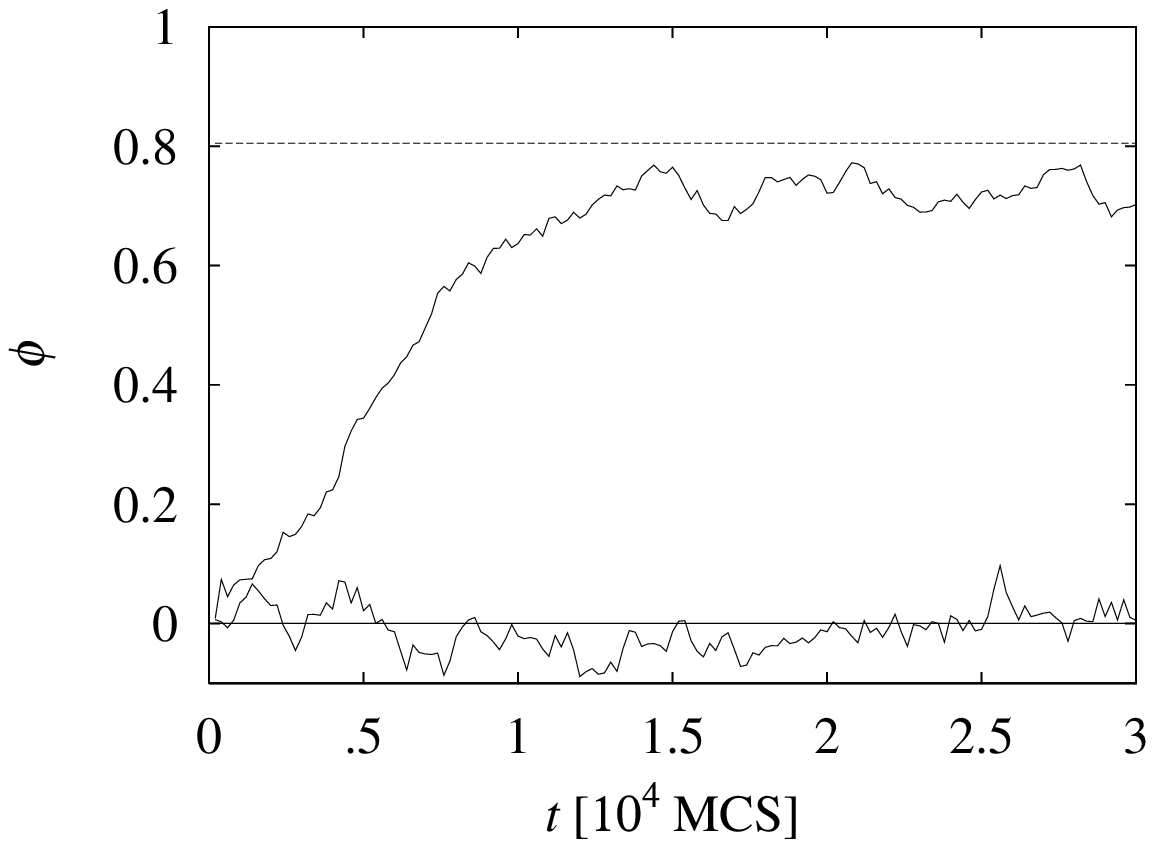}\\
(d) 
\end{center} 
\caption{Configurations of (a) A, (b) R, and (c) S molecules
with a concentration  $c_0=0.15$ 
at $3 \times 10^4$th MCS  with diffusion. 
(d) Evolution of the enanthiometric excess $\phi$.
Upper curve corresponds to the case with diffusions,
and the lower curve to that without diffusion, shown in Fig.3.
}
\label{fig4}
\end{figure}

\noindent
{\it Results and analysis for the case with diffusion}

Diffusion drastically changes the above situation at low densities.
Figures 4(a), (b) and (c), show configurations of A, R and S molecules, 
respectively, at
the concentration $c_0=0.15$ with diffusion and nonlinear autocatalysis
at the time $3 \times 10^4$th MCS.
The parameter values are $D=1/4,~ k_0=\lambda=10^{-3},~k_1=0,~
k_2=100k_0$, and the system size is $L^2 = 100^2$.
The R molecule in Fig. 4(b) increases their number 
at the cost of A and S molecules in Fig.4(a) and (c).
The ee approaches to the saturation value
$\phi_{\infty} \approx 0.73$ very quickly.
At the fixed point of the rate equation (6), the approximate value of the ee,
$\phi$, 
is given by eq.(7) as 0.871, quite larger than the simulation result.
However, since $k_2c_0^2$ is comparable to $k_0$, 
the asymptotic form (7) is no more valid
and the exact value is calculated to be 0.806, 
closer to the simulation result.
We have simulated larger system with sizes $L^2=200^2$ and $L^2=400^2$
with $k_2=100k_0$,
and found that the ee becomes non zero as well, though for a large system   
the initial incubation period with vanishing $\phi$ 
remains very long in some cases.
It indicates that there is a certain critical size of coherence for the chiral
symmetry breaking to take place, and the diffusion kinetics controls the 
propagation of symmetry breaking through the system.

With a linear autocatalysis and diffusion, our simulation shows that
the chiral symmetry will not be broken in the diluted system. 
Therefore, both
the nonlinearly autotalytic reaction together with recycling
and the diffusion
seem to be necessary for the chiral symmetry breaking in a dilute solution.

As we lower the concentration, the value of the ee decreases, and
at concentrations lower than a critical value about $c_{c} \approx 0.12$,
the system cannot sustain the state with broken chiral symmetry.
This is also expected from 
the rate equations (6): it has no symmetry-broken 
solution below the critical concentration
\begin{align}
c_{\tiny c} = \left( 2+ \frac{\lambda}{2k_0} \right)
\sqrt{\frac{k_0}{6k_2}},
\end{align}
which is $c_{\tiny c} =0.102$ for the present choice of the parameters,
$k_2=100k_0=100 \lambda$.

In fact, this is more clearly understood from the time evolution
 of the ee, derived from the rate equation (6), as
\begin{align}
\frac{d \phi}{dt} = \left[ 3k_2(c_0-a)^2(1- \phi^2 )-2k_0
\right] \frac{a}{c_0-a}  \phi .
\end{align}
Here $a$ is the time-dependent concentration of A molecule. 
The term proportional to $k_2$ in eq. (9)
represents that the nonlinear autocatalysis
amplifies the chiral symmetry breaking, whereas the term proportional to
$k_0$ suppresses the chiral symmetry breaking by the random 
and independent production of enantiomers, R and S.
The state with chiral symmetry looses its stability when the
coefficient of the linear term in $\phi$ in the right-hand side of eq.(9) 
is positive, and the state with a finite ee, $\phi$, can emerge.
Since $c_0-a$ represents the total concentration of R and S molecules which
is close to the initial concentration $c_0$, the nonlinear symmetry breaking
effect becomes weak for a dilute system, and below the critical concentration
the random creation of racemics dominates.

As for the critical concentration, there seems to be a discrepancy
between the simulation result and the rate equation analysis.
Since the rate equation corresponds to the mean field approximation
without fluctuation, the critical concentration in the simulation
might turn out to be a little larger than the theoretical prediction.
Another possibility is the finiteness of the
diffusion constant. In the rate equation, we assume a homogeneous situation,
corresponding to an infinitely fast diffusion. With a finite diffusion,
the system is influenced by the spatial fluctuation. 
There are also many other possibilities; finite simulation time and size
in the Monte Carlo simulation, etc.
More studies are required on the critical behaviors of this dynamical
phase transition.

\noindent
{\it Summary}

We have proposed a simple lattice model of chemical reaction with
molecular diffusion, and studied the chirality selection.
The nonlinear autocatalysis is shown to be indispensable for the
selection. In a diluted solution, molecular motion such as diffusion 
is necessary to accomplish the selection.
In nature, molecules are in water and the convection should provide 
much more efficient molecular movement.
If the initial concentration of the reactant $c_0$ is too low, 
below the critical concentration,
one can produce only racemic mixture of R and S.
The critical concentration depends on the ratio of the non-autocatalytic
to the nonlinearly autocatalytic rate coefficients, $k_0/k_2$.
The asymptotic ee value $\phi$ differs from the complete $\phi=1$
by a factor proportional to $k_0/k_2$.
If the initial production of R or S molecules from the reactant A
is triggered by minute external effect, $k_0$ might be very small
and the almost complete homochirality be achieved.

In the whole analysis, the back reaction from the chiral products, R and S, to
the reactant A is always assumed. Without it, the reaction stops before
attaining the full selection, since the reactant A is consumed up.
The recycling is necessary to develope the selection.
But it should be smaller than the critical value
$\lambda_c =2 k_0 (c_0 \sqrt{6k_2/k_0}-2)$ in order that the
symmetry-broken states exist.
As $\lambda$ is much too small, the system takes a long relaxation time to
reach the broken-symmetry state. 
There seems to be an appropriate range of values of $\lambda$ to
achieve the chiral symmetry breaking.



\begin{thebibliography}{99}

\bibitem{bonner88}
W. A. Bonner: Topics Stereochem. {\bf 18}  (1988) 1.

\bibitem{feringa+99}
B. L. Feringa and R. A. van Delden: Angew. Chem. Int. Ed. {\bf 38} (1999)
3418.

\bibitem{mason+85}
S. F. Mason and G. E. Tranter: Proc. R. Soc. Lond. {\bf A 397} (1985) 45.

\bibitem{kondepudi+85}
D. K. Kondepudi and G. W. Nelson: Nature {\bf 314} (1985) 438.

\bibitem{meiring87}
W. J. Meiring: Nature {\bf 329} (1987) 712.

\bibitem{bada95}
J. L. Bada: Nature {\bf 374} (1995) 594.

\bibitem{bailey+98}
J. Bailey, A. Chrysostomou, J. H. Hough, T. M. Gledhill, A. McCall, S. Clark, 
F. M\'enard and M. Tamura:
Science {\bf 281} (1998) 672.

\bibitem{hazen+01}
R. M. Hazen, T. R. Filley and G. A. Goodfriend: Proc. Natl. Acad. Sci. 
{\bf 98} (2001) 5487.

\bibitem{frank53}
F. C. Frank: Biochimi. Biophys. Acta {\bf 11}
 (1953) 459.

\bibitem{soai+90}
K. Soai, S. Niwa and H. Hori: J. Chem. Soc. Chem. Commun.  (1990) 982.

\bibitem{soai+95}
K. Soai, T. Shibata, H. Morioka and K. Choji: Nature {\bf 378} (1995) 767.

\bibitem{sato+01}
I. Sato, D. Omiya, K. Tsukiyama, Y. Ogi and K. Soai: 
Tetrahedron Asymmetry {\bf 12} (2001) 1965.


\bibitem{sato+03}
I. Sato, D. Omiya, H. Igarashi, K. Kato, Y. Ogi, K. Tsukiyama and K. Soai: 
Tetrahedron Asymmetry {\bf 14} (2003) 975.

\bibitem{kondepudi+01}
D. K. Kondepudi and K. Asakura, Acc. Chem. Res. {\bf 34} (2001) 946.

\bibitem{sh+04}
Y. Saito and H. Hyuga, J. Phys. Soc. Jpn {\bf 73} (2004) 33.



\bibitem{stauffer85}
D. Stauffer, {\it Introduction to percolation theory}
, (Taylor and Francis, London, 1985).


\end{thebibliography}
\end{document}